\documentclass[aps,prl,onecolumn,groupedaddress]{revtex4-1}
\usepackage{graphicx}
\usepackage{here}
\usepackage{xspace}
\usepackage{amsmath}
\usepackage{amssymb}
\begin{document}
\title{Three superconducting phases with different categories of pairing in hole- and electron-doped LaFeAs$_{1-x}$P$_x$O}

\author{S.~Miyasaka}
\affiliation{Department of Physics, Osaka University, Osaka 560-0043, Japan.}
\author{M.~Uekubo}
\affiliation{Department of Physics, Osaka University, Osaka 560-0043, Japan.}
\author{H.~Tsuji}
\affiliation{Department of Physics, Osaka University, Osaka 560-0043, Japan.}
\author{M.~Nakajima}
\affiliation{Department of Physics, Osaka University, Osaka 560-0043, Japan.}
\author{T.~Shiota}
\affiliation{Graduate School of Engineering Science, Osaka University, Osaka 560-8531, Japan.}
\author{H.~Mukuda}
\affiliation{Graduate School of Engineering Science, Osaka University, Osaka 560-8531, Japan.}
\author{S.~Tajima}
\affiliation{Department of Physics, Osaka University, Osaka 560-0043, Japan.}
\author{H.~Sagayama}
\affiliation{Condensed Matter Research Center and Photon Factory, Institute of Materials Structure Science, High Energy Accelerator Research Organization, Tsukuba 305-0801, Japan.}
\author{H.~Nakao}
\affiliation{Condensed Matter Research Center and Photon Factory, Institute of Materials Structure Science, High Energy Accelerator Research Organization, Tsukuba 305-0801, Japan.}
\author{R.~Kumai}
\affiliation{Condensed Matter Research Center and Photon Factory, Institute of Materials Structure Science, High Energy Accelerator Research Organization, Tsukuba 305-0801, Japan.}
\author{Y.~Murakami}
\affiliation{Condensed Matter Research Center and Photon Factory, Institute of Materials Structure Science, High Energy Accelerator Research Organization, Tsukuba 305-0801, Japan.}

\date{\today}

\draft

\begin{abstract}
The phase diagram of LaFeAs$_{1-x}$P$_x$O system has been extensively studied through hole- and electron-doping as well as As/P-substitution. 
It has been revealed that there are three different superconducting phases with different Fermi surface (FS) topologies and thus with possibly different pairing glues. 
One of them is well understood as spin fluctuation-mediated superconductivity within a FS nesting scenario. 
Another one with the FSs in a bad nesting condition must be explained in a different context such as orbital or spin fluctuation in strongly correlated electronic system. 
In both phases, $T$-linear resistivity was commonly observed when the superconducting transition temperature $T_{\rm c}$ becomes the highest value, indicating that the strength of bosonic fluctuation determines $T_{\rm c}$. 
In the last superconducting phase, the nesting condition of FSs and the related bosonic fluctuation are moderate. 
Variety of phase diagram characterizes the multiple orbital nature of the iron-based superconductors which are just near the boundary between weak and strong correlation regimes.

\end{abstract}


\maketitle
\section{I. INTRODUCTION}
In spite of a lot of experimental and theoretical studies on iron-based superconductors, their superconductivity mechanism has not been clarified yet. 
Just after the discovery of these superconducting compounds~\cite{Kamihara}, it has been reported that the system can be described within a moderate electron correlation regime where the band theory works well. 
Then, antiferromagnetic (AFM) spin fluctuation due to Fermi surface (FS) nesting was considered as a strong candidate for the pairing interaction of superconductivity~\cite{Kuroki,Mazin}. 

Recently, however, there have been many experimental reports that cannot be understood in this framework based on the hole and electron FS nesting. 
One of the typical examples is FeSe system where the FS near the $\Gamma$ point is missing~\cite{Qian,Zhang,He,Miyata} and thus the hole and electron FS nesting scenario is impossible. 
The strong electron correlation regime is necessary to understand this system~\cite{FeSe}. 
Another important issue is the contribution of orbital degree of freedom~\cite{Kontani}. 
The electronic nematicity observed in various iron-based superconductors suggests that the orbital fluctuation might play a crucial role in the electronic properties of these compounds~\cite{Yoshizawa,Kuo}. 
In mother materials of iron-based superconductors, the strong interplay of spin, charge, orbital and lattice degrees of freedom causes the spin and orbital orders concomitantly with structural phase transition~\cite{Fernandes}. 
For example, our recent work on the lattice dynamics of SrFe$_2$As$_2$ has revealed that a strong anisotropy in phonon dispersion results from the magneto-elastic coupling via a particular orbital~\cite{Murai}. 

In order to specify the mechanism of superconductivity, it is important to find the parameter which is correlated with the superconducting transition temperature ($T_{\rm c}$). 
In the case of iron-based superconductors, it is empirically known that the pnictogen ($Pn$) height from the Fe-plane and/or the $Pn$-Fe-$Pn$ bond angle ($\alpha$) shows a clear correlation with $T_{\rm c}$~\cite{Mizoguchi,Lee}. 
Using various compounds whose structural parameters are intentionally changed, it is possible to study systematically the electronic state in relation to superconductivity. 
One of the good platforms for such a study is $Ln$FeAsO ($Ln$: rare earth element) system. 
Hereafter we call it 1111-system. 
In our previous studies, we have investigated the As/P substitution effects on the electronic state of LaFeAs$_{1-x}$P$_x$O$_{1-y}$F$_y$ for $y=$0, 0.05 and 0.10~\cite{Miyasaka,Lai}. 
In the present work, we have extended this study, covering a wider composition range. 
This fully systematic study has revealed three distinct superconducting phases in different categories.

\section{II. EXPERIMENTS}
Choosing LaFeAsO as a parent compound, we can dope holes into it by substituting Sr for La, and electrons by substituting F or H for oxygen~\cite{Kamihara,Mu,Wen,Lu,Iimura,Hosono,Hiraishi,Kobayashi}. 
Polycrystalline samples of hole doped La$_{0.9}$Sr$_{0.1}$FeAs$_{1-x}$P$_x$O and electron doped LaFeAs$_{1-x}$P$_x$O$_{0.86}$F$_{0.14}$ with $x=$0-1.0 were synthesized by a solid state reaction method. 
A mixture of LaAs, LaP, Fe$_2$O$_3$, Fe, LaF$_3$, and SrAs powder with the stoichiometric ratio was pressed into a pellet in a pure Ar filled glove box and annealed at 1130 $^{\circ }$C for 40 hours in evacuated silica tubes. 
Heavily electron doping cannot be achieved by F-substitution but possible by H-substitution~\cite{Iimura,Hosono,Hiraishi,Kobayashi}. 
In order to synthesize H-substituted samples, it is necessary to use a high pressure furnace. 
Polycrystalline samples of electron doped LaFeAs$_{1-x}$P$_x$O$_{1-y}$H$_{y}$ ($x=$0-1.0, $y=$0.25 and 0.30) were synthesized under high pressures. 
A mixture of LaAs, LaP, Fe$_2$O$_3$, Fe, LaH$_2$ powder was pressed into a BN capsule and then heated at 1100 $^{\circ }$C for 2 hours under a pressure of 4GPa. 

All the samples were characterized by powder X-ray diffraction using Cu $K\alpha$ radiation at room temperature. 
Figure 1 shows the powder X-ray diffraction patterns for La$_{0.9}$Sr$_{0.1}$FeAs$_{1-x}$P$_x$O, LaFeAs$_{1-x}$P$_x$O$_{0.86}$F$_{0.14}$ and LaFeAs$_{1-x}$P$_x$O$_{0.70}$H$_{0.30}$. 
The diffraction pattern for LaFeAs$_{1-x}$P$_x$O$_{1-y}$H$_{y}$ with $y=$0.25 is very similar to that for $y=$0.30, and we only show the results of $y=$0.30 in Fig. 1. 
Almost all the diffraction peaks in Fig. 1 can be assigned to the calculated Bragg peaks for the tetragonal $P$4/$nmm$ symmetry. 
The in-plane ($a$) and out-of-plane lattice constants ($c$) were obtained by the least squares fitting of the X-ray diffraction data. 

\begin{figure}[htp]
\begin{center}
\includegraphics[width=0.7\textwidth]{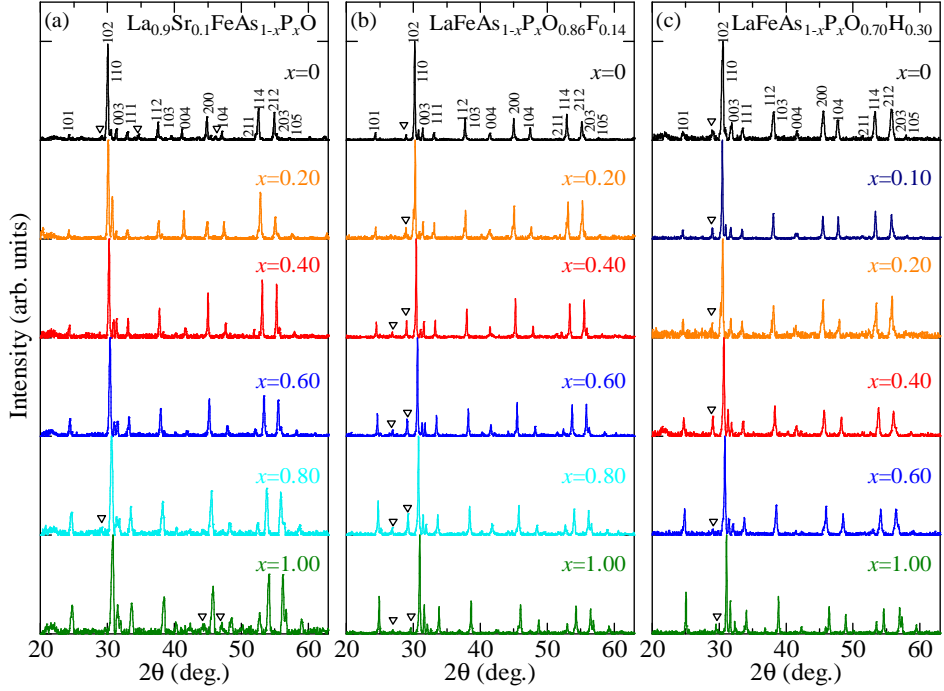}\\
\end{center}
\caption{(Color online) Powder X-ray diffraction patterns 
for (a) La$_{0.9}$Sr$_{0.1}$FeAs$_{1-x}$P$_x$O, (b) LaFeAs$_{1-x}$P$_x$O$_{0.86}$F$_{0.14}$ and (c) LaFeAs$_{1-x}$P$_x$O$_{0.70}$H$_{0.30}$ with various $x$s at room temperature. 
Almost all the diffraction peaks are indexed 
assuming the tetragonal structure with the $P$4/$nmm$ symmetry. 
The peaks indicated by triangles are due to impurities. 
}
\label{fig1}
\end{figure}

The crystal structures of LaFeAs$_{1-x}$P$_x$O$_{0.86}$F$_{0.14}$ and LaFeAs$_{1-x}$P$_x$O$_{1-y}$H$_{y}$ ($y=$0.25 and 0.30) were characterized by high resolution powder X-ray diffraction with the X-ray beam energy of 11.5 keV or 15 keV at room temperature at BL-8A/8B of Photon Factory in KEK, Japan. 
The lattice constants, pnictogen heights and $Pn$-Fe-$Pn$ bond angles were calculated from the measured data by Rietveld analysis~\cite{Izumi}. 
The lattice constants estimated by Rietveld analysis are almost the same as those determined from powder X-ray diffraction data using Cu $K\alpha$ radiation. 
La$_{0.9}$Sr$_{0.1}$FeAs$_{1-x}$P$_x$O samples were easily decomposed in air, so we could not perform the powder X-ray diffraction experiments for these samples using synchrotron X-ray beam. 

Electrical resistivity was measured by a standard four-probe method. 
$T_{\rm c}$ of almost all samples was determined by zero resistivity. 
In La$_{0.9}$Sr$_{0.1}$FeAs$_{1-x}$P$_x$O, the onset temperature of resistive 
transition is also shown. 
Magnetic susceptibility measurements were performed with SQUID magnetometer 
in an applied field of 10 Oe. 
The Knight shift and the nuclear spin-lattice relaxation rate (1/$T_1$) of $^{31}$P-NMR were measured at the field of $\sim$ 11.93 T. 

\section{III. RESULTS}

\begin{figure}[htp]
\begin{center}
\includegraphics[width=0.7\textwidth]{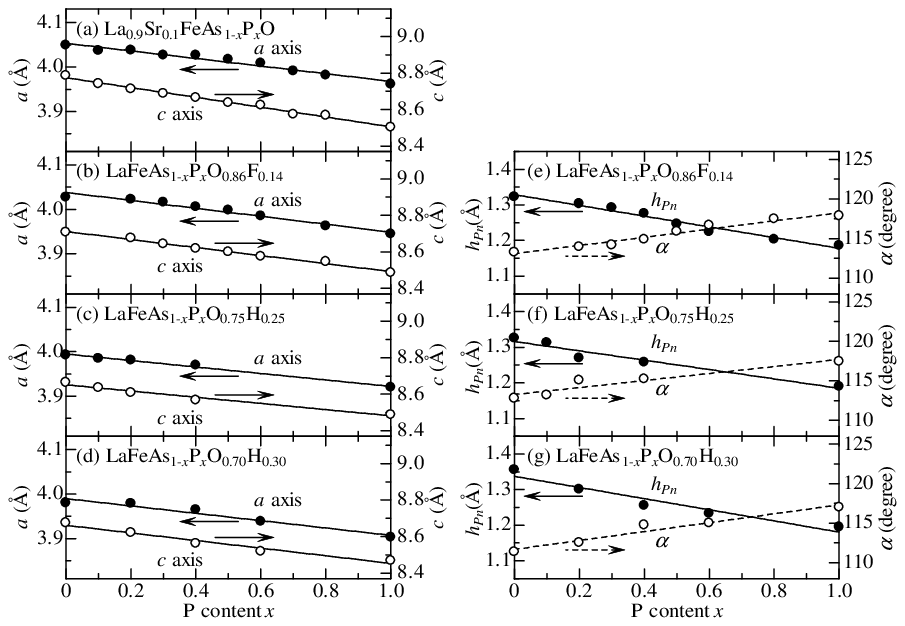}\\
\end{center}
\caption{The $x$ dependence of the lattice constants $a$ and $c$ of (a) La$_{0.9}$Sr$_{0.1}$FeAs$_{1-x}$P$_x$O, (b) LaFeAs$_{1-x}$P$_x$O$_{0.86}$F$_{0.14}$, (c) LaFeAs$_{1-x}$P$_x$O$_{0.75}$H$_{0.25}$ and (d) LaFeAs$_{1-x}$P$_x$O$_{0.70}$H$_{0.30}$ at room temperature. The $x$ dependence of the pnictogen height from the Fe plane $h_{Pn}$ and As/P-Fe-As/P bond angle $\alpha$ of (e) LaFeAs$_{1-x}$P$_x$O$_{0.86}$F$_{0.14}$, (f) LaFeAs$_{1-x}$P$_x$O$_{0.75}$H$_{0.25}$ and (g) LaFeAs$_{1-x}$P$_x$O$_{0.70}$H$_{0.30}$.}
\label{fig2}
\end{figure}

In Figs. 2(a)-(d) are plotted the lattice constants $a$ and $c$ as a function of P content $x$ for La$_{0.9}$Sr$_{0.1}$FeAs$_{1-x}$P$_x$O, LaFeAs$_{1-x}$P$_x$O$_{0.86}$F$_{0.14}$, LaFeAs$_{1-x}$P$_x$O$_{0.75}$H$_{0.25}$ and LaFeAs$_{1-x}$P$_x$O$_{0.70}$H$_{0.30}$. 
The lattice constant values for $x=$0 (As 100$\%$) in Figs. 2(a)-(d) are consistent with those in the previous reports~\cite{Mu,Wen,Hosono}. 
In all the systems, the lattice constants linearly decrease with $x$, following the Vegard's law. 
This proves that As/P solid solution compounds were successfully synthesized. 
The primary effect of P-substitution is the shrinkage of the lattice because the ionic radius of P is smaller than that of As. 

The pnictogen height from the Fe plane $h_{Pn}$ and As/P-Fe-As/P bond angle $\alpha$ were estimated for LaFeAs$_{1-x}$P$_x$O$_{0.86}$F$_{0.14}$, LaFeAs$_{1-x}$P$_x$O$_{0.75}$H$_{0.25}$ and LaFeAs$_{1-x}$P$_x$O$_{0.70}$H$_{0.30}$, as shown in Figs. 2(e)-(g), respectively. 
The values of $h_{Pn}$ and $\alpha$ for $x=$0 (As 100$\%$) are consistent with the previous work by Hosono and Matsuishi~\cite{Hosono}. 
These quantities also linearly change with $x$. 
The increase of $\alpha$ with $x$ implies that P substitution does not uniformly shrink the lattice but deforms the Fe-As/P tetrahedron. 
This deformation of the Fe-As/P tetrahedron is considered to play a crucial role in the non-monotonic changes of many physical properties that were observed in our previous studies~\cite{Miyasaka,Lai}. 
It is expected that the effect of tetrahedron deformation is different in different carrier doping levels. 
We show the results for each doping region. 

\begin{figure}[htp]
\begin{center}
\includegraphics[width=0.45\textwidth]{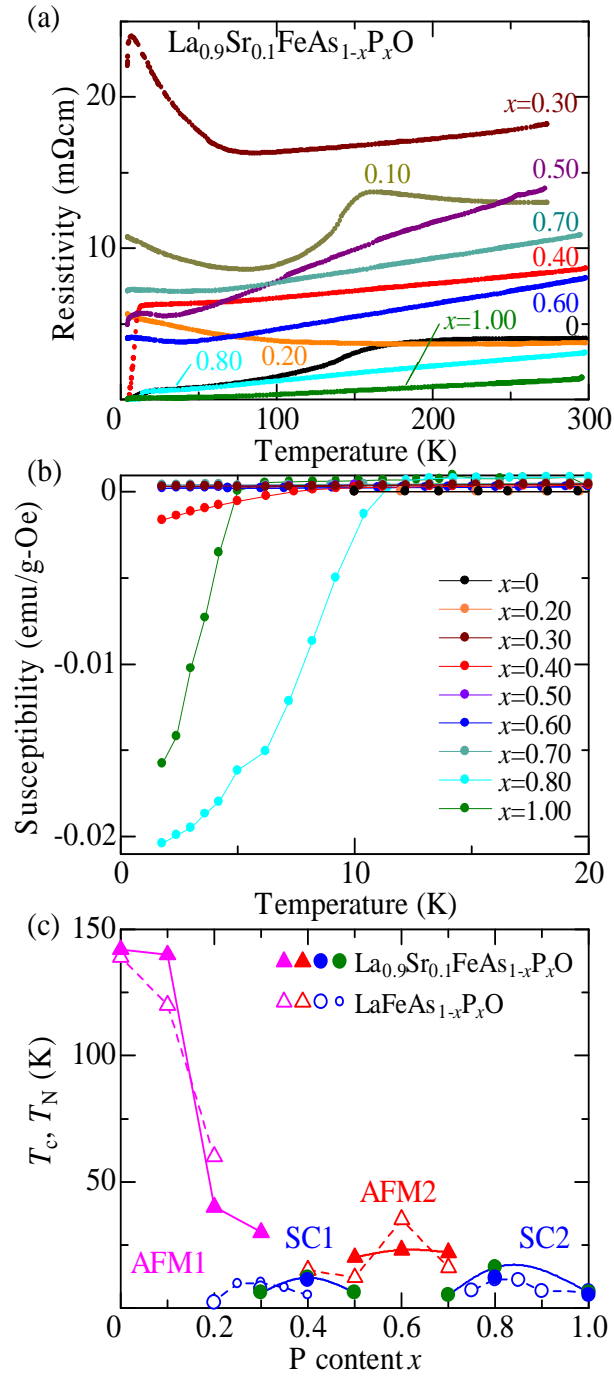}\\
\end{center}
\caption{(Color online) (a) The temperature dependence of the electrical resistivity and (b) magnetic susceptibility of La$_{0.9}$Sr$_{0.1}$FeAs$_{1-x}$P$_x$O with various $x$s. (c) The P content ($x$) dependence of critical temperature $T_{\rm c}$ and N$\acute{\text{e}}$el temperature $T_N$ of La$_{0.9}$Sr$_{0.1}$FeAs$_{1-x}$P$_x$O shown by closed symbols. The blue and green circles indicate $T_{\rm c}$ determined by zero resistivity and by the onset of resistive transition, respectively. $T_{\rm c}$ and $T_N$ of LaFeAs$_{1-x}$P$_x$O in previous reports are also plotted in panel (c) by large open symbols~\cite{Lai} and small ones~\cite{Wang,Kitagawa}. Solid and broken lines are the guide for eyes.}
\label{fig3}
\end{figure}

First, we show the results of hole doped system, La$_{0.9}$Sr$_{0.1}$FeAs$_{1-x}$P$_x$O. The temperature ($T$) dependences of resistivity ($\rho$) and magnetic susceptibility ($\chi$) of La$_{0.9}$Sr$_{0.1}$FeAs$_{1-x}$P$_x$O are shown for various P contents in Figs. 3(a) and (b), respectively. 
The composition dependence of resistivity behavior is complicated. 
At $x=$1.0 (P 100$\%$) resistivity is very low and exhibits a metallic $T$-dependence. 
Superconductivity is observed around 5 K. 
When P concentration is reduced, resistivity slightly increases and $T_{\rm c}$ increases up to $\sim$12 K at $x=$0.8. 
With further decreasing P content, superconductivity disappears and $\rho$($T$) shows a kink at around 20 K. 
We consider that this kink feature corresponds to the AFM phase transition accompanied with the structural phase transition~\cite{Klauss}. 
At lower P contents, the kink of $\rho$($T$) disappears and the superconductivity manifests itself again, for example, at 12 K at $x=$0.4. 
However, this superconducting composition region is very narrow and the other AFM phase appears below $x=$0.3. 
The superconductivity for $x=$0.4, 0.8 and $x=$1.0 was also confirmed by magnetic susceptibility. 

Figure 3(c) illustrates the phase diagram of La$_{0.9}$Sr$_{0.1}$FeAs$_{1-x}$P$_x$O, $T_{\rm c}$ and N$\acute{\text{e}}$el temperature $T_N$ being plotted as a function of $x$. 
We can clearly see two superconducting phases (SC1 and SC2) and 
two AFM phases (AFM1 and AFM2). 
This is very similar to the phase diagram of LaFeAs$_{1-x}$P$_x$O~\cite{Lai,Wang,Kitagawa} that is indicated in the same figure by open symbols and dashed curves. 
Although in LaFeAs$_{1-x}$P$_x$O the AFM state between $x=$0.4 and 0.7 was observed only by nuclear magnetic resonance (NMR) but not clearly seen in $\rho$($T$)~\cite{Lai}, it is visible in $\rho$($T$) for the case of La$_{0.9}$Sr$_{0.1}$FeAs$_{1-x}$P$_x$O. 
Therefore, we can safely conclude that there exists the second AFM phase (AFM2) between the two superconducting phases in the parent and hole doped compounds. 
The effect of hole doping is the expansion of the first AFM phase (AFM1) near $x=$0, and the shift of the first superconducting phase (SC1) to the higher $x$ region. 

In this work, the synchrotron X-ray diffraction could not be performed at low temperatures for La$_{0.9}$Sr$_{0.1}$FeAs$_{1-x}$P$_x$O because the samples are easily decomposed. To confirm the structural phase transition together with the pahse transition to AFM2 state in La$_{0.9}$Sr$_{0.1}$FeAs$_{1-x}$P$_x$O, a furthe diffraction study is required. 

\begin{figure}[htp]
\begin{center}
\includegraphics[width=0.45\textwidth]{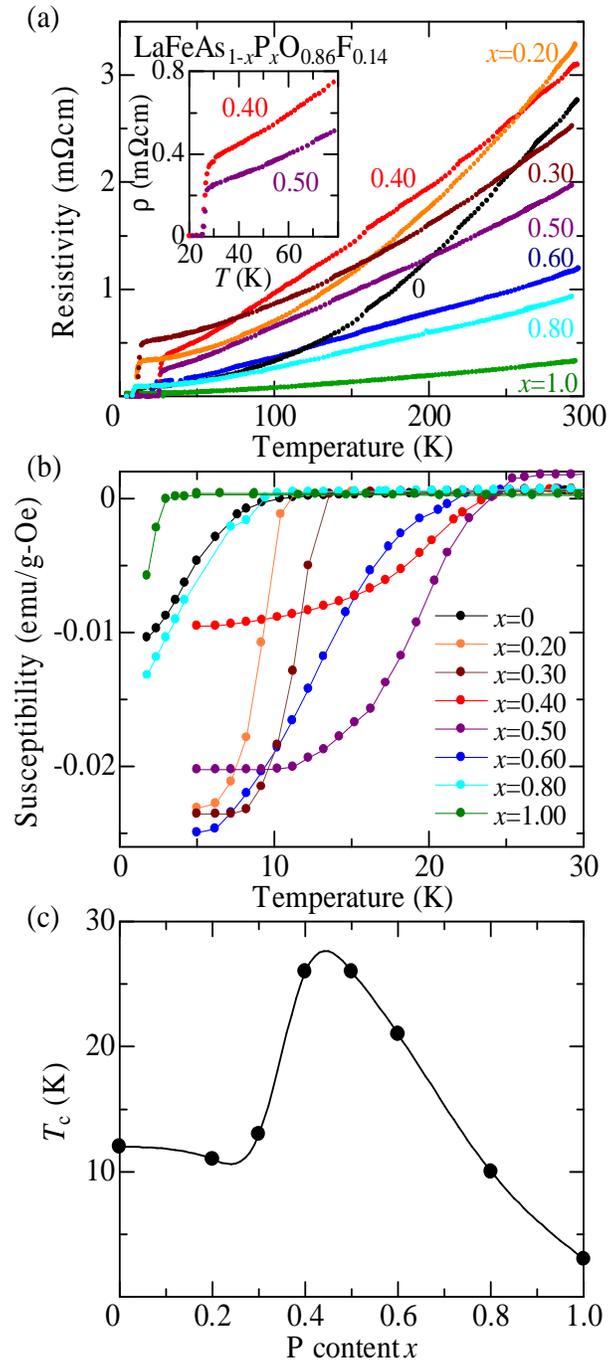}\\
\end{center}
\caption{(Color online) (a) The temperature dependence of the electrical resistivity and (b) magnetic susceptibility of LaFeAs$_{1-x}$P$_x$O$_{0.86}$F$_{0.14}$ with various $x$s. (c) The P content ($x$) dependence of critical temperature $T_{\rm c}$ of LaFeAs$_{1-x}$P$_x$O$_{0.86}$F$_{0.14}$. Solid line is the guide for eyes. The inset of top panel (a) shows the temperature ($T$) dependence of the electrical resistivity $\rho$ for $x=$0.40 and 0.50 at low temperatures.}
\label{fig4}
\end{figure}

Next, the results for electron doped system, LaFeAs$_{1-x}$P$_x$O$_{0.86}$F$_{0.14}$, are presented in Fig. 4. 
Figure 4(a) shows the $T$ dependence of resistivity for various $x$ values. 
Roughly speaking, resistivity does not change with $x$ so much. 
All the samples show a metallic $T$-dependence of resistivity with a superconducting transition. 
Bulk superconductivity was also confirmed by magnetic susceptibility, as shown in Fig. 4(b). 
The resistivity is the lowest at $x=$1.0, which is commonly observed in all the studied As/P substitution systems. 
As shown in Fig. 1(a) and the inset, the $T$-dependence becomes almost linear, $\rho$($T$)$\sim T$, and $T_{\rm c}$ shows a maximum value at $x=$0.4-0.5. 
At the lower and the higher P contents, $T$-dependence approaches the Fermi liquid behavior $\rho \sim T^2$. 
The $T_{\rm c}$ is plotted as a function of $x$ in Fig. 4(c). 
In spite of the linear change of all the structure parameters (Figs. 2(b) and (e)), $T_{\rm c}$ changes non-monotonically with $x$, peaking at $x=$0.4-0.5. 

\begin{figure}[htp]
\begin{center}
\includegraphics[width=0.45\textwidth]{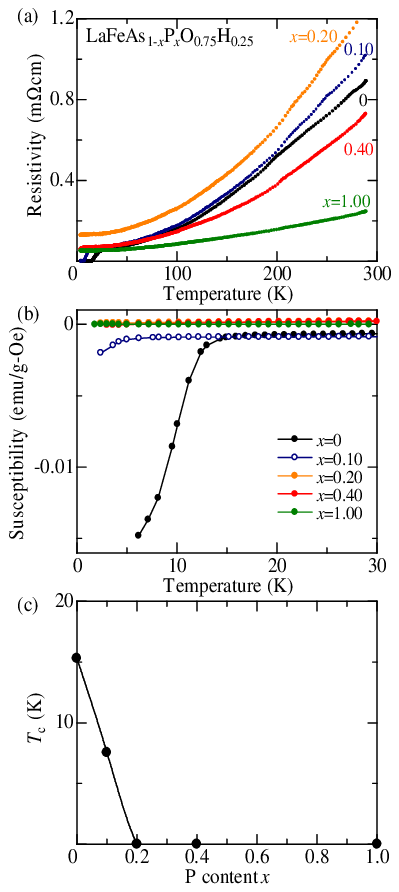}\\
\end{center}
\caption{(Color online) (a) The temperature dependence of the electrical resistivity and (b) magnetic susceptibility of LaFeAs$_{1-x}$P$_x$O$_{0.75}$H$_{0.25}$ with various $x$s. (c) The P content ($x$) dependence of critical temperature $T_{\rm c}$ of LaFeAs$_{1-x}$P$_x$O$_{0.75}$H$_{0.25}$. Solid line is the guide for eyes.}
\label{fig5}
\end{figure}

\begin{figure}[htp]
\begin{center}
\includegraphics[width=0.45\textwidth]{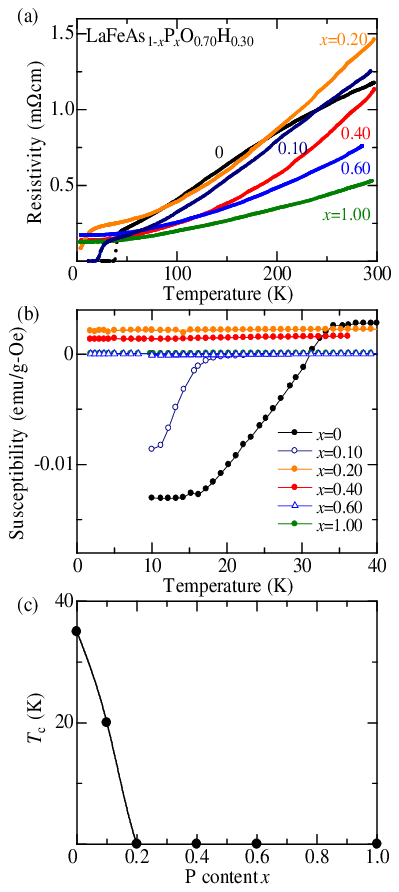}\\
\end{center}
\caption{(Color online) (a) The temperature dependence of the electrical resistivity and (b) magnetic susceptibility of LaFeAs$_{1-x}$P$_x$O$_{0.70}$H$_{0.30}$ with various $x$s. (c) The P content ($x$) dependence of critical temperature $T_{\rm c}$ of LaFeAs$_{1-x}$P$_x$O$_{0.70}$H$_{0.30}$. Solid line is the guide for eyes.}
\label{fig6}
\end{figure}

Finally the results of heavily electron doped systems, LaFeAs$_{1-x}$P$_x$O$_{0.75}$H$_{0.25}$ and LaFeAs$_{1-x}$P$_x$O$_{0.70}$H$_{0.30}$, are examined. 
By using hydrogen, we could achieve the heavy electron-doping in 1111-system. 
The $T$-dependences of resistivity for all the P compositions are shown in Figs. 5(a) and 6(a). 
As is clearly seen, all the samples show metallic $\rho$($T$) with very small resistivity values. 
For example, the residual resistivity is less than 250 $\mu \Omega$cm for all the compositions although the samples are polycrystals. 
As shown in Figs. 5(a), 5(b), 6(a), and 6(b), the superconductivity appears only around $x=$0 in these systems. 
The maximum $T_{\rm c}$($\sim$35 K) is observed at $x=$0 (As 100$\%$) for $y=$0.30. 
This value is higher than those for lower doping levels ($y$). 
The $T_{\rm c}$ value for $y=$0.25 is much lower than the case of $y=$0.30. 

When $x$ is slightly increased, $T_{\rm c}$ is rapidly suppressed and eventually disappears above $x=$0.2. 
The $T_{\rm c}$ values of both H-doped systems are plotted as a function of $x$ in Figs. 5(c) and 6(c). 
It should be noted that the non-superconducting samples with $x>$0.2 are not insulators but very good metals, exhibiting $T^2$-resistivity. 
This suggests that there is no AFM phase in this compositional region.

\section{IV. DISCUSSION}

\begin{figure}[htp]
\begin{center}
\includegraphics[width=0.7\textwidth]{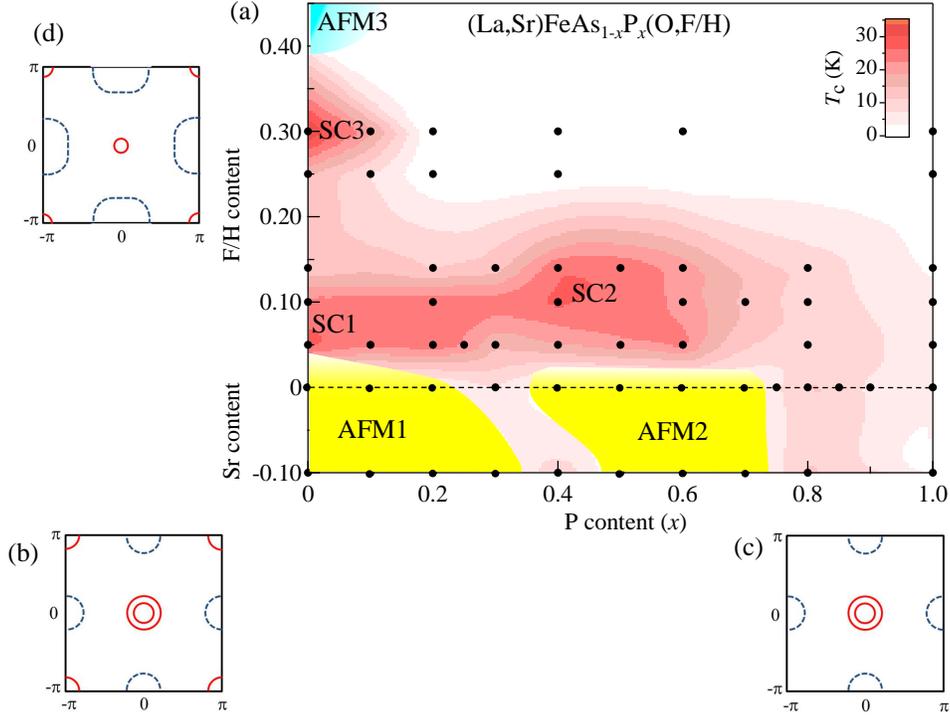}\\
\end{center}
\caption{(Color online) (a) The schematic phase diagram of (La,Sr)FeAs$_{1-x}$P$_x$(O,F/H). Three blue areas indicate the antiferromagnetic phases (AFM1, AFM2 and AFM3). The contour lines of $T_{\rm c}$ are drawn by extrapolation on the basis of the data shown in Figs. 3(c), 4(c), 5(c), 6(c) and previous results~\cite{Miyasaka,Lai}. The dots indicate the position of materials that were actually synthsized. The schematic Fermi surfaces for (b) LaFeAsO, (c) LaFePO and (d) heavily electron-doped LaFeAsO cases in the unfolded Brillouin zone. The solid red and broken blue lines indicate the hole and electron Fermi surfaces, respectively.}
\label{fig7}
\end{figure}

All the data of $T_{\rm c}$ and $T_N$ are summarized in Fig. 7(a), together with the results for F-contents of $y=$0, 0.05 and 0.10 in our previous works~\cite{Miyasaka,Lai}. 
Here we can identify three superconducting regions. 
The first one (SC1) is near the AFM1 phase, while the second one (SC2) is close to the AFM2 phase. 
The last one (SC3) is located at very high H-doping compositions. 

According to the band calculation~\cite{Kuroki}, the Fermi surfaces of LaFeAsO are expected as illustrated in Fig. 7(b). 
Here the nesting between hole and electron FSs is crucial for spin fluctuation. 
Since the sizes of hole and electron FSs are comparable, the nesting condition is relatively good, which causes AFM order in AFM1. 
When holes are doped into the parent compound LaAsFeO, the hole FSs near the $\Gamma$-point slightly expand while the electron FS near ($\pm \pi$, 0) or (0, $\pm \pi$) shrinks. 
This makes the nesting condition better, resulting in the stability of AFM1 phase. 
This could be the reason for the expansion of AFM1 phase with Sr-substitution.

For LaFePO, the band calculation predicts that the $d_{xy}$ FS near ($\pi$,$\pi$) is missing~\cite{Kuroki,Usui}. (See Fig. 7(c).)
Therefore, the nesting between $d_{yz}$ and $d_{zx}$ FS is important for spin fluctuation. 
The AFM2 phase is considered to be in a strong limit of this fluctuation. 
When electrons are doped, this spin correlation is weakened and the AFM2 phase vanishes. 
Instead, the SC2 phase emerges. 
Therefore, it is likely that the superconductivity in the SC2 phase is mediated by the two dimensional spin fluctuation via the $d_{yz}$/$d_{xz}$ FS nesting. 

\begin{figure}[htp]
\begin{center}
\includegraphics[width=0.45\textwidth]{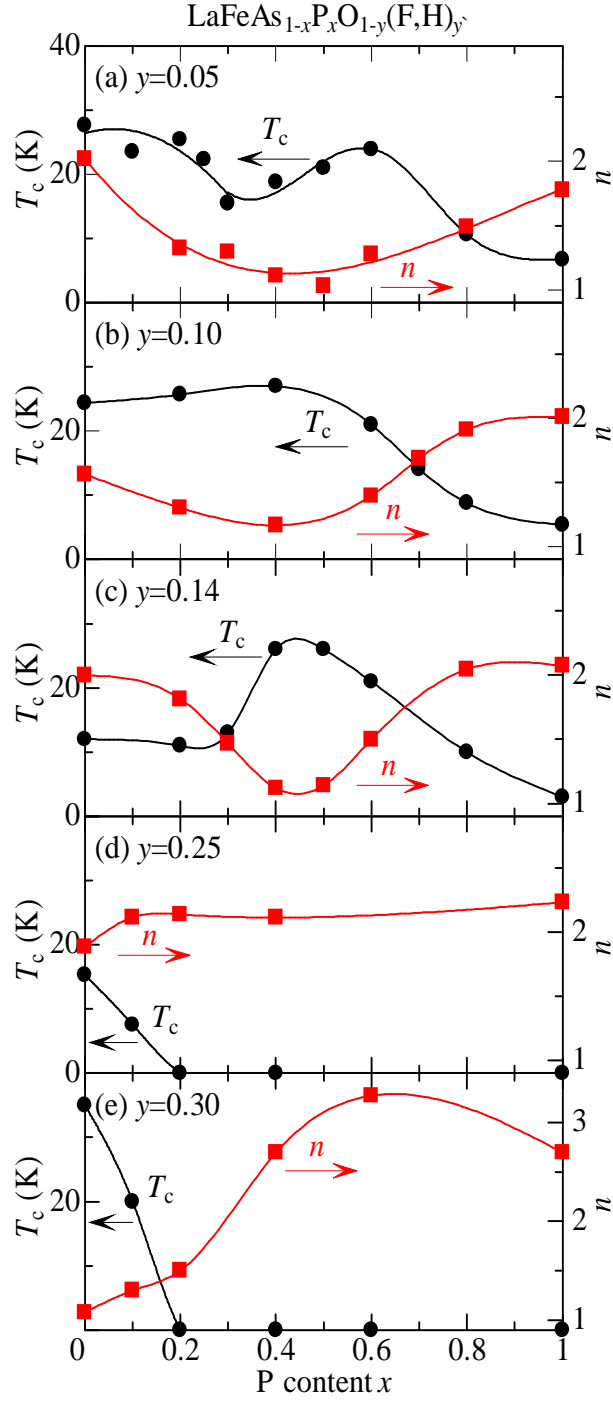}\\
\end{center}
\caption{(Color online) The P concentration ($x$) dependence of critical temperature $T_{\rm c}$ (black circles) and the exponent $n$ in $\rho$($T$)$= \rho_0+AT^n$ (red squares) for (a) LaFeAs$_{1-x}$P$_x$O$_{1-y}$F$_{y}$ with $y=$0.05~\cite{Lai}, (b) $y=$0.10~\cite{Miyasaka}, (c) $y=$0.14, (d) LaFeAs$_{1-x}$P$_x$O$_{1-y}$H$_{y}$ with $y=$0.25 and (e) $y=$0.30, respectively. The solid lines are the guide for eyes. }
\label{fig8}
\end{figure}

The strength of spin fluctuation can be monitored by the power ($n$) of $T$ in resistivity $\rho$($T$)$= \rho_0 + AT^n$. 
The $x$-dependence of $n$ is summarized in Fig. 8. 
The fitting of $\rho$($T$) was performed between 100 K and the onset temperature of resistive transition. 
The fitting parameters of $\rho$($T$) for $x=$0.40, 0.50 in LaFeAs$_{1-x}$P$_x$O$_{0.86}$F$_{0.14}$, and $x=$0 in LaFeAs$_{1-x}$P$_x$O$_{0.70}$H$_{0.30}$, where $\rho$($T$)$\sim T$ as presented in Figs. 4(a) and 6(a), are shown as examples in the Appendix. 
In the electron doped systems with $y=$0.05, 0.10 and 0.14, $n$ decreases from 2 (at $x=$1.0) to 1 with decreasing $x$. 
At the $x$-composition where $n \sim$1, $T_{\rm c}$ reaches a maximum value. 
This can be understood as follows. 
At $x=$1.0, spin fluctuation is weak and the system is close to the Fermi liquid. 
As $x$ decreases, spin fluctuation strength increases and the system turns to a non-Fermi liquid state near the quantum critical point, where the $T$-linear resistivity and the $T_{\rm c}$ maximum are observed. 

\begin{figure}[htp]
\begin{center}
\includegraphics[width=0.45\textwidth]{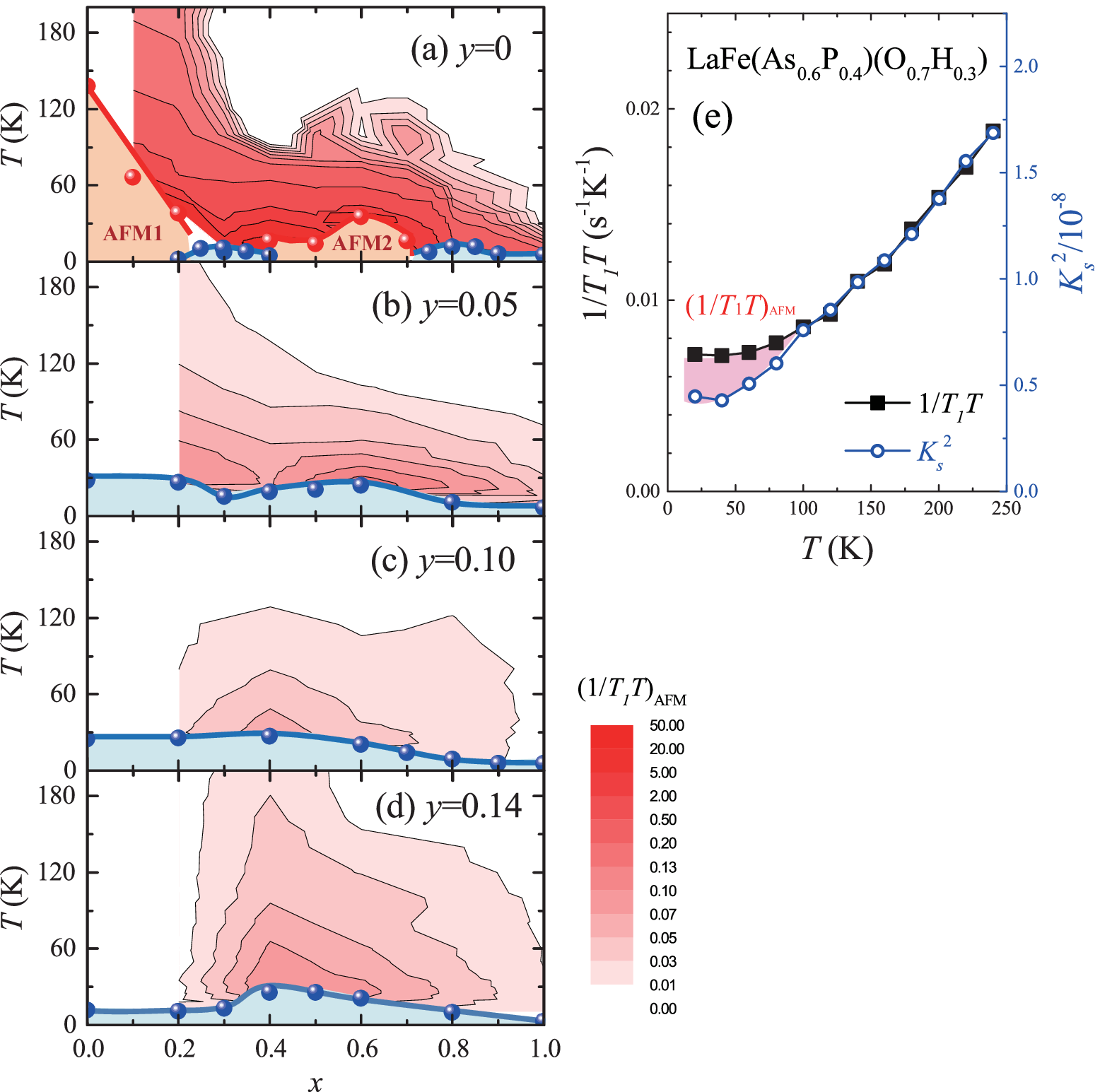}\\
\end{center}
\caption{(Color online) The contour plots of (1/$T_1T$)$_{AFM}$ for (a) $y=$0, (b) $y=$0.05, (c) $y=$0.10 and (d) $y=$0.14 of LaFeAs$_{1-x}$P$_x$O$_{1-y}$F$_{y}$~\cite{Shiota}. (e) Temperature ($T$) dependence of 1/$T_1T$ (closed squares) and $K_s^2$ (open circles) for LaFeAs$_{1-x}$P$_x$O$_{1-y}$H$_{y}$ with $x=$0.40 and $y=$0.30. The hatched area is denoted as (1/$T_1T$)$_{AFM}$.}
\label{fig9}
\end{figure}

The $T$-linear resistivity is caused by some bosonic fluctuation but not necessarily by spin fluctuation. 
However, in the case of SC2, the corresponding AFM spin fluctuation was clearly observed by nuclear magnetic resonance (NMR) experiments. 
Figures 9(a)-(d) are contour plots of the AFM fluctuation component of 1/$T_1T$ for various compositions, where $T_1$ is a nuclear spin relaxation time estimated from the $^{31}$P-NMR experiment. 
Here we assume that (1/$T_1T$) is decomposed as (1/$T_1T$)$=$(1/$T_1T$)$_{AFM}+$(1/$T_1T$)$_0$~\cite{Shiota}. 
The first term represents the strength of AFM fluctuation due to FS nesting, while the second one is $q$-independent and proportional to the square of density of states at $E_F$. 
The second term was estimated by assuming that its $T$-dependence is the same as that of $K_s^2$. 
Here, $K_s$ is the spin part of Knight shift, which is proportional to the density of states at $E_F$. 
Figure 9 clearly demonstrates that the composition ($x$) region of the strong AFM fluctuation corresponds to the $x$-region of high $T_{\rm c}$ for $y=$0.05, 0.10, and 0.14. It means that when the AFM fluctuation is enhanced, the $T_{\rm c}$ is raised in the SC2 phase. 

What is remarkable here is that even if the system approaches the quantum critical point with increasing As concentration it does not go into an AFM ordered phase but into another superconducting phase (SC1). 
As discussed in our previous paper, this is linked to the band crossover with As substitution, namely, the sinking down of $d_{z^2}$ band below $E_F$ and the rising up of the $d_{xy}$ band above $E_F$~\cite{Miyasaka,Lai}. 

Next, we discuss the SC3 phase. 
In the heavily electron doped state, the hole FS must be radically shrunk, while the electron FS is expanded~\cite{Suzuki}. (See Fig. 7(d).) 
The imbalance of FS sizes makes the nesting condition worse. 
In this circumstance, it is inappropriate to take the nesting scenario under the assumption of moderately electron correlation. 
Actually, the NMR measurement confirms that the low energy spin fluctuation is very weak in LaFeAs$_{0.6}$P$_{0.4}$O$_{0.7}$H$_{0.3}$. 
As shown in Fig. 9(e), the increase of the 1/$T_1T$ due to AFM spin fluctuations, evaluated as $|$(1/$T_1T$)$_{AFM}| \sim$0.002, is significantly smaller than $|$(1/$T_1T$)$_{AFM}| \sim$0.07 at ($x$, $y$)$=$(0.40, 0.10) that corresponds to the highest $T_{\rm c}$ composition in SC2 phase. 
Note that the $T$-linear resistivity with high $T_{\rm c}$ is observed at $x=$0 for $y=$0.30 (Fig. 8(e)), while the lower-$T_{\rm c}$ samples for $y=$0.25 and the non-superconducting ones for $y=$0.25 and 0.30 show $T^2$- or larger power law behavior of $\rho$($T$). 
It means that a strong bosonic fluctuation governs the physical properties in the SC3 phase. 

One of the candidates for this bosonic fluctuation is the spin fluctuation due to the next nearest neighbor hopping of $d_{xy}$ orbital electrons~\cite{Usui,Suzuki}. 
This is based on a strong electron correlation regime in real space, in contrast to the FS nesting picture in $k$-space. 
According to this scenario, the rapid suppression of $T_{\rm c}$ with increasing $x$ can be explained by the disappearance of $d_{xy}$ FS near ($\pi$,$\pi$), because this orbital mainly contributes to the spin fluctuation via the next nearest neighbor hopping. 
Since the system rapidly changes to the Fermi liquid with increasing $x$, electron correlation strongly depends on the orbitals forming FSs. 
Another possibility is superconductivity mediated by other bosonic fluctuation such as orbital fluctuation. 
The contribution of orbital degree of freedom has been widely discussed in the iron-based superconductors~\cite{Kontani}. 
The electronic nematicity could be a smoking gun of orbital fluctuation~\cite{Yoshizawa,Kuo,Fernandes}.

Finally, we return to the discussion of SC1 phase. In this phase, it has been found that the spin fluctuation monitored by 1/$T_1T$ of NMR is not strongly enhanced at low temperatures~\cite{Mukuda}. 
The $T$-dependence of resistivity also suggests a weaker bosonic fluctuation~\cite{Miyasaka,Lai}. 
The expected FSs are illustrated in Fig. 7(b). 
Here the nesting condition is moderate, as pointed out in refs. \cite{Usui,Suzuki}. 
Usui and coworkers proposed that the spin fluctuation due to FS nesting (low energy limit) does not play a major role but the spin fluctuation due to finite energy hopping contributes to superconductivity. 
Actually the 1/$T_1T$ of NMR indicated that AFM spin fluctuation starts to develop at higher temperature in SC1 than in SC2, which may indicate the contribution of spin fluctuation with higher energy scale~\cite{Shiota}. 
Because of the multiple-orbital character, various kinds of spin fluctuations with various energy scales are possibly involved in the superconductivity. 
Another possibility is the contribution of other bosonic fluctuation. 
To specify the pairing mechanism in this phase, further study is required.

\section{V. CONCLUSION}
We have systematically studied the As/P substitution effects on the electronic state of LaFeAs$_{1-x}$P$_x$O for various carrier doping levels ($y$). 
In the hole-doped La$_{0.9}$Sr$_{0.1}$FeAs$_{1-x}$P$_x$O, the temperature-dependent resistivity indicates that there exist two AFM phases (AFM1 and AFM2) and two superconducting phases (SC1 and SC2) as in LaFeAs$_{1-x}$P$_x$O. 
In electron-doped LaFeAs$_{1-x}$P$_x$O with the doping level of $y=$0.14, SC1 state is destabilized by heavy electron doping, while SC2 survives around $x=$0.40-0.50. 
In further electron-doped LaFeAs$_{1-x}$P$_x$O$_{1-y}$H$_y$ with $y=$0.30, aother superconducting phase (SC3) appears at $x \sim$0. 
In these regions of ($x$, $y$)$=$(0.40-0.50, 0.14) and (0, 0.30), resistivity shows non Fermi liquid behaviors, indicating the existence of strong bosonic fluctuation. 

The extensive study over a wide $x$-$y$ composition range has revealed three distinct superconducting phases with different FS topologies. 
One of them (SC2) can be well understood as the superconductivity mediated by AFM spin fluctuation via the nesting of $d_{yz}$/$d_{zx}$ FSs, because $T_{\rm c}$ is clearly correlated with the spin fluctuation strength monitored by 1/$T_1T$ in NMR and the power $n$ in $\rho \sim T^n$. 
The superconducting phase (SC3) in heavily electron doped region should be considered in a strong correlation regime where the FS nesting scenario does not work. 
In this phase, a particular orbital ($d_{xy}$) seems to play a crucial role in superconductivity. 
In the last superconducting phase (SC1), the nesting condition is moderate and the bosonic fluctuation monitored by $\rho$($T$) and 1/$T_1T$ is also moderate. 
Contribution of spin fluctuation with various energy scales is one of the possible candidates for pairing glue in SC1, although the possibility of other bosonic fluctuation cannot be excluded.

\

\begin{acknowledgments}
We thank K. Kuroki and H. Usui for helpful discussion. 
This work was supported by Grants-in-Aid for Scientific Research from MEXT and JSPS, and by the JST project (TRIP and IRON-SEA) in Japan. 
This work was performed under the approval of the Photon Factory Program Advisory Committee (Proposal No. 2012S2-005, 2014S2-001, 2016G145 and 2017G044).

\end{acknowledgments}

\section{VI. APPENDIX}

\begin{table}[h]
\caption{\label{tab:table1}%
The fitting parameters, $\rho_0$, $A$ and $n$ of temperature-dependent resistivity ($\rho$($T$)$=\rho_0 + AT^n$) for ($x$, $y$)$=$(0.40, 0.14) and (0.50, 0.14) in LaFeAs$_{1-x}$P$_x$O$_{1-y}$F$_{y}$, and ($x$, $y$)$=$(0, 0.30) in LaFeAs$_{1-x}$P$_x$O$_{1-y}$H$_{y}$. }
\begin{ruledtabular}
\begin{tabular}{cccc}
($x$, $y$) & $\rho_0$ (m$\Omega$ cm) & $A$ ($\mu \Omega$ cm/K$^n$) & $n$\\
\colrule
(0.4, 0.14) & 0.16$\pm$0.02 & 4.8$\pm$0.9 & 1.12$\pm$0.05\\
(0.5, 0.14) & 0.10$\pm$0.01 & 2.8$\pm$0.7 & 1.14$\pm$0.05\\
(0, 0.30) & 0.05$\pm$0.01 & 2.4$\pm$0.6 & 1.08$\pm$0.05\\
\end{tabular}
\end{ruledtabular}
\end{table}

Here we provide the fitting parameters of temperature($T$)-dependent resistivity $\rho$($T$) for the optimal $T_{\rm c}$ samples ($x=$0.40, 0.50 in LaFeAs$_{1-x}$P$_x$O$_{0.86}$F$_{0.14}$, and $x=$0 in LaFeAs$_{1-x}$P$_x$O$_{0.70}$H$_{0.30}$) to supplement the main text. 
As shown in Figs. 4(a) and 6(a), the $\rho$($T$) can be expressed as $\rho$($T$)$=\rho_0 + AT^n$ at low temperatures, where $\rho_0$ is residual resistivity, $n$ the power of $T$ and $A$ the coefficient. The fitting of $\rho$($T$) was performed between 100 K and the onset $T$ of resistive transition. 
Table I shows the fitting results of $\rho_0$, $A$ and $n$ for $x=$0.40, 0.50 in LaFeAs$_{1-x}$P$_x$O$_{0.86}$F$_{0.14}$, and $x=$0 in LaFeAs$_{1-x}$P$_x$O$_{0.70}$H$_{0.30}$.




\begin{thebibliography}{99}
\bibitem{FeSe}In FeSe system, the nesting between the electron pockets still remains, and it cannot be ruled out as a pairing channel.
\end{thebibliography}
\end{document}